\documentclass[a4paper]{article}
\usepackage[utf8]{inputenc}
\usepackage[T1]{polski}
\usepackage{graphicx}\graphicspath{{img/}}
\usepackage{lmodern,fancyhdr,secdot,float,amsmath,amsfonts,amsthm,amssymb}
\usepackage{booktabs, tabularx}
\usepackage{enumerate}
\usepackage[margin=2cm]{geometry}
\usepackage{multicol}
\usepackage[hidelinks,unicode]{hyperref}
\usepackage{titlesec}
\usepackage{cite}
\usepackage{textcomp}
\usepackage{stfloats}
\usepackage{url}
\usepackage{verbatim}
\usepackage{multirow}
\usepackage[table,xcdraw]{xcolor}
\usepackage{colortbl}
\usepackage{algorithmic}
\usepackage[ruled,vlined]{algorithm2e}
\usepackage{array}
\usepackage[caption=false,font=normalsize,labelfont=sf,textfont=sf]{subfig}
\usepackage[normalem]{ulem}
\useunder{\uline}{\ul}{}
\titleformat{\section}{\centering\normalsize\bf}{\thesection.}{.5em}{\MakeUppercase}
\titleformat*{\subsection}{\bf\normalsize\selectfont}
\titleformat*{\subsubsection}{\bf\normalsize\selectfont}
\newcommand{\titlePL}[1]{\large\textbf{ #1}}
\newcommand{\titleEN}[1]{\normalsize #1}
\newcommand{\keywordsPL}[1]{\small\textbf{Słowa kluczowe:} #1}
\newcommand{\keywordsEN}[1]{\small\textbf{Keywords:} #1}
\newcommand{\abstractPL}[1]{\small\textbf{Streszczenie:} #1}
\newcommand{\abstractEN}[1]{\small\textbf{Abstract:} #1}

\definecolor{logo_color}{RGB}{40, 69, 166}

\begin{document}\thispagestyle{empty}\pagestyle{fancy}
\begin{minipage}[t]{0.5\textwidth}\vspace{0pt}%
\includegraphics[scale=0.9]{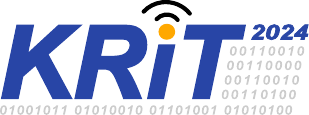}
\end{minipage}
\begin{minipage}[t]{0.45\textwidth}\vspace{12pt}%
\centering
\color{logo_color} KONFERENCJA RADIOKOMUNIKACJI\\ I TELEINFORMATYKI\\ KRiT 2024
\end{minipage}

\vspace{1cm}

\begin{center}
\titlePL{Propagacja sygnału radiowego w systemach 5G wyposażonych w~matryce IPR}

\titleEN{Radio signal propagation in 5G systems equipped with RISs}\medskip

Adam Samorzewski$^{1}$;
Adrian Kliks$^{2,3}$;

\medskip

\begin{minipage}[t]{1\textwidth}
\centering
\small $^{1}$ Politechnika Poznańska, Poznań, Polska, \href{mailto:email}{adam.samorzewski@doctorate.put.poznan.pl}\\
\small $^{2}$ Politechnika Poznańska, Poznań, Polska, \href{mailto:email}{adrian.kliks@put.poznan.pl}\\
\small $^{3}$ Uniwersytet Techniczny w Luleå, Luleå, Szwecja, \href{mailto:email}{adrian.kliks@ltu.se}\\
\end{minipage}

\medskip

\end{center}

\medskip

\begin{multicols}{2}
\noindent
\abstractPL{
W niniejszym artykule przeanalizowano charakterystykę propagacji sygnału radiowego na obszarze zawierającym się w granicach miasta Poznania (Polska). W rozważaniach przyjęto wykorzystanie sieci dostępu radiowego RAN ({\it ang. Radio Access Network}) systemu bezprzewodowego $5$. generacji ($5$G NR \textbf{--} {\it ang. New Radio}), w skład którego wchodzi $8$ stacji bazowych (SB) wykorzystujących technologię antenową SISO ({\it ang. Single Input Single Output}) lub MIMO ({\it ang. Multiple Input Multiple Output}) w zależności od przyjętej konfiguracji komórek sieciowych. Na badanym obszarze umieszczono również $15$~matryc odbijających zwanych Inteligentnymi Powierzchniami Rekonfigurowalnymi (IPR) oraz uwzględniono ich wpływ na propagację sygnału radiowego dla różnych wysokości ich zawieszenia.}
\medskip

\noindent
\abstractEN{
In this paper, the characteristics of radio signal propagation within the boundaries of the city of Poznan (Poland) are analyzed. The study considers the use of a Radio Access Network (RAN) of the $5$th generation wireless system ($5$G NR \textbf{--} New Radio), which includes $8$~base stations (BSs) utilizing Single Input Single Output (SISO) or Multiple Input Multiple Output (MIMO) antenna technology depending on the adopted configuration of network cells. Additionally, $15$ reflecting arrays known as Reconfigurable Intelligent Surfaces (RISs) were placed in the studied area, and their impact on radio signal propagation at different suspension heights was taken into account.\footnote{Copyright © 2024 SIGMA-NOT. Personal use is permitted. For any other purposes, permission must be obtained from the SIGMA-NOT by emailing sekretariat@sigma-not.pl. This is the author’s version of an article that has been published in the journal entitled \textit{Telecommunication Review -- Telecommunication News} (PL: \textit{Przegląd Telekomunikacyjny -- Wiadomości Telekomunikacyjne}) by the SIGMA-NOT. Changes were made to this version by the publisher before publication, the final version of the record is available at: https://dx.doi.org/10.15199/59.2024.4.61. To cite the paper use: A. Samorzewski, A.~Kliks, “Radio signal propagation in 5G systems equipped with RISs" (PL: "Propagacja sygnału radiowego w systemach 5G wyposażonych w matryce IPR"), \textit{Telecommunication Review -- Telecommunication News} (PL: \textit{Przegląd Telekomunikacyjny -- Wiadomości Telekomunikacyjne}), 2024, no. 4, pp.~280--283, doi: 10.15199/59.2024.4.61 or visit https://sigma-not.pl/publikacja-150600-2024-4.html.}}
\medskip

\noindent
\keywordsPL{5G, Inteligentne Powierzchnie Rekonfigurowalne, propagacja sygnału radiowego, tłumienie ścieżki, systemy bezprzewodowe.}
\medskip

\noindent
\keywordsEN{5G, path loss, radio signal propagation, Reconfigurable Intelligent Surfaces, wireless systems.}

\section{Wstęp}
\label{section:introduction}
Obecnie stadium rozważań dotyczących zarówno teraźniejszych $\left(5\text{G}\right)$, jak i przyszłych $\left(6\text{G}\right)$ generacji systemów mobilnych wprowadza do literatury naukowej coraz to nowsze rozwiązania propagacji sygnału. W ostatnich latach duże zainteresowanie zyskały tzw. Inteligentne Powierzchnie Rekonfigurowalne ({\it ang. Reconfigurable Intelligent Surfaces} -- RISs), które cechują się zdolnością odbijania sygnału radiowego i kierunkowania go w stronę konkretnego obszaru. Taka właściwość miałaby umożliwić mobilnym operatorom ({\it ang. Mobile Network Operators} -- MNOs) manipulowanie pokryciem radiowym w celu {\it doświetlania} terenów, na których terminale odbiorcze mają utrudniony dostęp do usług mobilnych. W dodatku wykorzystanie matryc IPR może skutecznie doprowadzić do redukcji interferencji międzysygnałowych i/lub narażenia użytkowników na działanie pola elektromagnetycznego (PEM). Wykorzystanie wspomnianych matryc IPR mogłoby znaleźć swoje szczególne zastosowanie w obrębie terenów miejskich, gdzie nadawany sygnał radiowy jest załamywany i rozpraszany przez liczne przeszkody znajdujące się na jego drodze. Niemniej jednak na terenach wiejskich aktywne matryce IPR (tj. takie, które dodatkowo wzmacniają przekierowywany sygnał radiowy) mogą zostać wdrożone jako przekaźniki, redukując przy tym liczbę stacji bazowych potrzebnych do pokrycia rozważanego obszaru \cite{Huang, Tang, SamorzewskiJTIT2023, SamorzewskiKRiT2023, SamorzewskiSoftCOM2023}.

W niniejszym artykule przeanalizowano wykorzystanie wspomnianych matryc IPR na obszarze miejskim badając przy tym ich wpływ na charakterystykę tłumienia ścieżki ({\it ang. Path Loss} -- PL) propagowanego sygnału radiowego. W rozważaniach wzięto również pod uwagę wpływ zawieszenia inteligentnych matryc na pokrycie badanego terenu. 

Artykuł został podzielony na następujące części: sekcja druga opisuje rozważany scenariusz systemowy; sekcja trzecia zawiera informacje dotyczące metodologii przeprowadzanych badań oraz konfiguracji symulacyjnych; treść sekcji czwartej składa się z wyników badań zaprezentowanych w formie zarówno grafów, jak i tablic ze średnimi wartościami obserwowanych parametrów systemowych; piąta i ostatnia sekcja podsumowuje całą pracę poprzez przedstawienie wniosków wyciągniętych na podstawie zaobserwowanych rezultatów.

\section{Rozważany scenariusz}
\label{section:scenario}
Rozważana sieć została umieszczona na terenie miasta Poznania \textbf{--} dokładnie wzięto pod uwagę Stary Rynek oraz ulice w jego sąsiedztwie \cite{SamorzewskiKRiT2023, SamorzewskiSoftCOM2023, SamorzewskiGLOBECOM2023}. W skład systemu mobilnego wchodzi $8$ stacji bazowych, gdzie każda z nich posiada $3$~komórki transmitujące sygnał radiowy na różnych częstotliwościach \textbf{--} $800$, $2100$ i $3500$ MHz. Dodatkowo każda z tych komórek wykorzystuje systemy antenowe transmitujące sygnał zgodnie z technikami SISO ($800$ i $2100$ MHz) lub MIMO ($3500$ MHz), które cechują się odmienną liczbą aktywnych elementów \textbf{--} $1$ (SISO) oraz $64$ (MIMO) przypadających na pojedynczy transciver \cite{Castellanos, SamorzewskiGLOBECOM2023}.

Ponadto na rozważanym terenie pomiędzy budynkami rozmieszczono $15$ pasywnych matryc IPR w taki sposób, że każda z nich jest w zasięgu przynajmniej jednej stacji bazowej. Matryce są umieszczone domyślnie na krawędzi każdego budynku, na którym się znajdują, jednakże ich wysokość może zostać zmanipulowana. 

Do zaprojektowania systemu mobilnego (które przekładało się m.in. na ustalenie lokalizacji stacji bazowych i budynków, jak i na wytyczenie wymiarów budowli znajdujących się na terenie systemu bezprzewodowego oraz określenie planowanego obszaru pokrycia sygnałem radiowym) wykorzystano dane rzeczywiste udostępnione w bazie danych BTSearch \cite{NetworkData} oraz w Systemie Informacji Przestrzennej Miasta Poznania \cite{AreaData}.

\section{Konfiguracja symulacyjna}
\label{section:simulation}
Badania zostały przeprowadzone w formie symulacyjnej przy użyciu dedykowanego oprogramowania o nazwie GRAND ({\it Green Radio Access Network Design}) \cite{Castellanos, SamorzewskiGLOBECOM2023}.

Głównym celem badań było przeanalizowanie charakterystyki tłumienia ścieżki sygnału radiowego na badanym terenie miejskim. Do ewaluacji wartości tłumienia ścieżki sygnału radiowego w danym punkcie przyjęto następujące modele propagacyjne: 
\begin{itemize}
    \item $3$GPP TR $38.901$ UMa ({\it ang. Urban Macro}) \textbf{--} propagacja sygnału LOS ({\it ang. Line-Of-Sight}) i NLOS ({\it ang. Non-Line-Of-Sight}) od stacji bazowej \cite{3GPP},
    \item RIS FFBC ({\it ang. RIS Far Field Beamforming Case}) \textbf{--} propagacja sygnału od stacji bazowej poprzez odbicie od matrycy IPR \cite{Tang}.
\end{itemize}

Co więcej, warunki propagacyjne zostały zaaranżowane (symulacyjnie) dla scenariusza systemowego, w~którym początkowo żadna z dostępnych matryc nie jest wykorzystywana do pośredniej propagacji sygnału radiowego, a następnie wszystkie nich są brane pod uwagę podczas estymacji charakterystyki tłumienia jego ścieżki. W~dodatku dla rozważanego scenariusza przetestowano kilka wariantów konfiguracji symulacyjnej, które różnią się między sobą, chociażby wysokością zawieszenia matryc odbijających sygnał radiowy. Poniżej wyszczególniono każdy ze wspomnianych przypadków wziętych pod uwagę podczas rozważań symulacyjnych:
\begin{enumerate}[I.]
    \item Matryce IPR zawieszone na domyślnej wysokości $h_\text{RIS}\text{ }\left[\text{m}\right]$ (tj. na wysokościach prawie równych wysokościom budynków, na których się znajdują).
    \item Matryce IPR zawieszone na wysokości $h_\text{RIS}+5\text{ }\left[\text{m}\right]$.
    \item Matryce IPR zawieszone na wysokości $h_\text{RIS}+10\text{ }\left[\text{m}\right]$.
\end{enumerate}

Wypunktowany powyżej zróżnicowany zbiór przypadków testowych pozwoli przeanalizować wpływ zarówno wysokości zawieszenia matryc IPR, jak i średniej wartości różnicy pomiędzy wysokościami zawieszenia układów antenowych stacji bazowych oraz matryc odbijających wiązkę radiową od tychże węzłów dostępowych (czyli de facto znajdujących się w ich zasięgu) na charakterystykę propagacyjną sygnału radiowego w obrębie rozważanego terenu pokrycia.

Konfiguracja parametrów symulacyjnych dla rozważanej sieci mobilnej została zawarta w kolejno zaprezentowanych tabelach (tj. Tab.~\ref{table:configuration_network_design} i \ref{table:configuration_ris_design}). Pierwsza z nich (Tab.~\ref{table:configuration_network_design}) zawiera wartości parametrów wykorzystanych do emulowania działania stacji bazowych w oprogramowaniu symulacyjnym. Wyszczególnione we wspomnianej tabeli parametry zawierają informacje dotyczące wykorzystywanego pasma, konfiguracji układów nadawczo-odbiorczych oraz modelowania warunków propagacyjnych dla każdego typu komórki sieciowej. Z kolei na Tab.~\ref{table:configuration_ris_design} składają się wartości dla parametrów symulacyjnych proponujących konkretnie zdefiniowany rodzaj inteligentnych matryc IPR. Wśród tych właściwości znajdują się takie cechy jak liczba oraz wymiary aktywnych elementów, wysokość zawieszenia matryc na budynkach, czy chociażby wykorzystany model propagacyjny do oszacowania potencjalnego tłumienia sygnału.

\begin{table}[H]
\centering
\caption{Konfiguracja parametrów (SB)
\cite{Castellanos, SamorzewskiGLOBECOM2023, 3GPP, Björnson, AreaData, NetworkData}
}
\label{table:configuration_network_design}
    \resizebox{0.49\textwidth}{!}{\begin{tabular}{|l|c|c|c|cccc|}
    \arrayrulecolor[HTML]{002060}\hline
    \multirow{3}{*}{\cellcolor[HTML]{002060}}   & \multirow{3}{*}{\cellcolor[HTML]{002060}{\color[HTML]{FFFFFF}}} & \multirow{3}{*}{\cellcolor[HTML]{002060}{\color[HTML]{FFFFFF} }} & \multirow{3}{*}{\cellcolor[HTML]{002060}{\color[HTML]{FFFFFF} }} & \multicolumn{3}{c|}{\cellcolor[HTML]{002060}{\color[HTML]{FFFFFF} Wartość}}                                                                                                                                        \\ \cline{5-7} 
                        \cellcolor[HTML]{002060}{\color[HTML]{FFFFFF}} & \cellcolor[HTML]{002060}{\color[HTML]{FFFFFF} Parametr}                           &  \cellcolor[HTML]{002060}{\color[HTML]{FFFFFF} Symbol}                     &  \cellcolor[HTML]{002060}{\color[HTML]{FFFFFF} Jednostka}                      & \multicolumn{3}{c|}{\cellcolor[HTML]{FFFFFF}{\color[HTML]{000000} \textit{Komórka stacji bazowej}}} \\ \cline{5-7}
                        \cellcolor[HTML]{002060}{\color[HTML]{FFFFFF}} & \cellcolor[HTML]{002060}{\color[HTML]{FFFFFF}}                           & \cellcolor[HTML]{002060}{\color[HTML]{FFFFFF}}                      & \cellcolor[HTML]{002060}{\color[HTML]{FFFFFF}}                      & \multicolumn{1}{c|}{$1\text{.}$}    & \multicolumn{1}{c|}{$2\text{.}$}   & \multicolumn{1}{c|}{$3\text{.}$} \\ \hline
    \multirow{1}{*}{\cellcolor[HTML]{89B0FF}{\color[HTML]{002060}}} 
                        \cellcolor[HTML]{89B0FF}{\color[HTML]{002060}} & \cellcolor[HTML]{E7E7E7}{\color[HTML]{000000} Liczba}                   & \cellcolor[HTML]{E7E7E7}{\color[HTML]{000000}$K_\text{BS}$}            & \cellcolor[HTML]{E7E7E7}{\color[HTML]{000000}\textbf{--}}                     &                                    \multicolumn{3}{c|}{\cellcolor[HTML]{E7E7E7}{\color[HTML]{000000}$8$}}      \\ \cline{2-7} 
                        \multirow{-2}{*}{\rotatebox[origin=c]{90}{\cellcolor[HTML]{89B0FF}{\color[HTML]{002060} Og.}}} & Technologia & \textbf{--} & \textbf{--} & \multicolumn{3}{c|}{$5\text{G}$} \\ 
                        \hline
    \multirow{1}{*}{\cellcolor[HTML]{89B0FF}{\color[HTML]{002060}}}
                        \cellcolor[HTML]{89B0FF}{\color[HTML]{002060}} & \cellcolor[HTML]{E7E7E7}{\color[HTML]{000000} Częstotliwość}                  & \cellcolor[HTML]{E7E7E7}{\color[HTML]{000000} $f$}            & \cellcolor[HTML]{E7E7E7}{\color[HTML]{000000}{$\left[\text{MHz}\right]$}}             & \multicolumn{1}{c|}{\cellcolor[HTML]{E7E7E7}{\color[HTML]{000000} $800$}}  & \multicolumn{1}{c|}{\cellcolor[HTML]{E7E7E7}{\color[HTML]{000000}$2100$}} & \multicolumn{1}{c|}{\cellcolor[HTML]{E7E7E7}{\color[HTML]{000000}$3500$}} \\ \cline{2-7} 
                        \cellcolor[HTML]{89B0FF}{\color[HTML]{002060}} & Szerokość pasma kanału          & $B_\text{w}$            & {$\left[\text{MHz}\right]$}      & \multicolumn{1}{c|}{$80$}   & \multicolumn{1}{c|}{$120$}  & \multicolumn{1}{c|}{$120$} \\ \cline{2-7} 
                        \cellcolor[HTML]{89B0FF}{\color[HTML]{002060}} & \cellcolor[HTML]{E7E7E7}{\color[HTML]{000000} L. użytych podnośne}           & \cellcolor[HTML]{E7E7E7}{\color[HTML]{000000} $N_\text{SC,u}$}            & \cellcolor[HTML]{E7E7E7}{\color[HTML]{000000} \textbf{--}}                     & \multicolumn{3}{c|}{\cellcolor[HTML]{E7E7E7}{\color[HTML]{000000} $320$}} \\ \cline{2-7} 
                        \cellcolor[HTML]{89B0FF}{\color[HTML]{002060}} & L. wszystkich podnośnych          & $N_\text{SC,t}$            & \textbf{--}                     & \multicolumn{3}{c|}{$512$} \\ \cline{2-7} 
                        \cellcolor[HTML]{89B0FF}{\color[HTML]{002060}} & \cellcolor[HTML]{E7E7E7}{\color[HTML]{000000} Współczynnik próbkowania}            & \cellcolor[HTML]{E7E7E7}{\color[HTML]{000000} SF}                    & \cellcolor[HTML]{E7E7E7}{\color[HTML]{000000} \textbf{--}}                     & \multicolumn{3}{c|}{\cellcolor[HTML]{E7E7E7}{\color[HTML]{000000} $1.536$}} \\ \cline{2-7} 
                        \cellcolor[HTML]{89B0FF}{\color[HTML]{002060}} & Współ. ponow. użycia pilota         & RF                    & \textbf{--}                     & \multicolumn{3}{c|}{$1$} \\ \cline{2-7} 
                        \cellcolor[HTML]{89B0FF}{\color[HTML]{002060}} & \cellcolor[HTML]{E7E7E7}{\color[HTML]{000000} Czas koherencji}             & \cellcolor[HTML]{E7E7E7}{\color[HTML]{000000} $t_\text{c}$}            & \cellcolor[HTML]{E7E7E7}{\color[HTML]{000000}{$\left[\text{ms}\right]$}}              & \multicolumn{3}{c|}{\cellcolor[HTML]{E7E7E7}{\color[HTML]{000000}$50$}} \\ \cline{2-7} 
                        \cellcolor[HTML]{89B0FF}{\color[HTML]{002060}} & Szerokość pasma koherencji        & $B_\text{c}$            & {$\left[\text{MHz}\right]$}             & \multicolumn{3}{c|}{$1$} \\ \cline{2-7}
                        \multirow{-9}{*}{\rotatebox[origin=c]{90}{\cellcolor[HTML]{89B0FF}{\color[HTML]{002060} Pasmo}}} & \cellcolor[HTML]{E7E7E7}{\color[HTML]{000000} Przestrzenny cykl pracy}         & \cellcolor[HTML]{E7E7E7}{\color[HTML]{000000} $S$}            & \cellcolor[HTML]{E7E7E7}{\color[HTML]{000000}{$\left[\text{\%}\right]$}}              & \multicolumn{1}{c|}{\cellcolor[HTML]{E7E7E7}{\color[HTML]{000000} $0$}}    & \multicolumn{1}{c|}{\cellcolor[HTML]{E7E7E7}{\color[HTML]{000000} $0$}}    & \multicolumn{1}{c|}{\cellcolor[HTML]{E7E7E7}{\color[HTML]{000000} $25$}} \\ \hline
    \multirow{1}{*}{\cellcolor[HTML]{89B0FF}{\color[HTML]{002060}}}  & Wysokość anteny             & $h_\text{BS}$            & {$\left[\text{m}\right]$}               & \multicolumn{3}{c|}{$\left(32, 46\right)$} \\ \cline{2-7} 
                        \cellcolor[HTML]{89B0FF}{\color[HTML]{002060}} & \cellcolor[HTML]{E7E7E7}{\color[HTML]{000000} L. elementów antenowych}           & \cellcolor[HTML]{E7E7E7}{\color[HTML]{000000} $M_\text{BS}$}            & \cellcolor[HTML]{E7E7E7}{\color[HTML]{000000} \textbf{--}}                     & \multicolumn{1}{c|}{\cellcolor[HTML]{E7E7E7}{\color[HTML]{000000} $1$}}    & \multicolumn{1}{c|}{\cellcolor[HTML]{E7E7E7}{\color[HTML]{000000} $1$}}    & \multicolumn{1}{c|}{\cellcolor[HTML]{E7E7E7}{\color[HTML]{000000} $64$}} \\ \cline{2-7} 
                        \cellcolor[HTML]{89B0FF}{\color[HTML]{002060}} & Zysk antenowy               & $G_\text{a}$            & {$\left[\text{dBi}\right]$}             & \multicolumn{1}{c|}{$16$}   & \multicolumn{1}{c|}{$18$}   & \multicolumn{1}{c|}{$24$} \\ \cline{2-7} 
                        \cellcolor[HTML]{89B0FF}{\color[HTML]{002060}} & \cellcolor[HTML]{E7E7E7}{\color[HTML]{000000} Tłumienie kabla antenowego}        & \cellcolor[HTML]{E7E7E7}{\color[HTML]{000000} $L_\text{f}$}            & \cellcolor[HTML]{E7E7E7}{\color[HTML]{000000}{$\left[\text{dBi}\right]$}}             & \multicolumn{1}{c|}{\cellcolor[HTML]{E7E7E7}{\color[HTML]{000000} $2$}}    & \multicolumn{1}{c|}{\cellcolor[HTML]{E7E7E7}{\color[HTML]{000000} $2$}}    & \multicolumn{1}{c|}{\cellcolor[HTML]{E7E7E7}{\color[HTML]{000000} $3$}} \\ \cline{2-7} 
                        \cellcolor[HTML]{89B0FF}{\color[HTML]{002060}} & Maks. moc nadawcza        & $P_\text{TX, max}$            & {$\left[\text{dBm}\right]$}             & \multicolumn{1}{c|}{$46$}   & \multicolumn{1}{c|}{$49$}   & \multicolumn{1}{c|}{$53$} \\ \cline{2-7} 
                        \multirow{-6}{*}{\rotatebox[origin=c]{90}{\cellcolor[HTML]{89B0FF}{\color[HTML]{002060} Transceivery}}} & \cellcolor[HTML]{E7E7E7}{\color[HTML]{000000} Współ. szumów}               & \cellcolor[HTML]{E7E7E7}{\color[HTML]{000000} NF}                    & \cellcolor[HTML]{E7E7E7}{\color[HTML]{000000}{$\left[\text{dB}\right]$}}              & \multicolumn{1}{c|}{\cellcolor[HTML]{E7E7E7}{\color[HTML]{000000} $8$}}    & \multicolumn{1}{c|}{\cellcolor[HTML]{E7E7E7}{\color[HTML]{000000} $8$}}    & \multicolumn{1}{c|}{\cellcolor[HTML]{E7E7E7}{\color[HTML]{000000} $7$}} \\ \hline
    \multirow{1}{*}{\cellcolor[HTML]{89B0FF}{\color[HTML]{002060}}}  & Model tł. ścieżki            & \textbf{--}                     & \textbf{--}                     & \multicolumn{3}{c|}{$3$GPP TR $38.901$ UMa \cite{3GPP}} \\ \cline{2-7} 
                        \cellcolor[HTML]{89B0FF}{\color[HTML]{002060}} & \cellcolor[HTML]{E7E7E7}{\color[HTML]{000000} Margines interferncji}        & \cellcolor[HTML]{E7E7E7}{\color[HTML]{000000} IM}                    & \cellcolor[HTML]{E7E7E7}{\color[HTML]{000000}{$\left[\text{dB}\right]$}}              & \multicolumn{3}{c|}{\cellcolor[HTML]{E7E7E7}{\color[HTML]{000000} $2$}}  \\ \cline{2-7} 
                        \cellcolor[HTML]{89B0FF}{\color[HTML]{002060}} & Margines Dopplera             & DM                    & {$\left[\text{dB}\right]$}              & \multicolumn{3}{c|}{$3$} \\ \cline{2-7} 
                        \cellcolor[HTML]{89B0FF}{\color[HTML]{002060}} & \cellcolor[HTML]{E7E7E7}{\color[HTML]{000000} Margines zaników}                & \cellcolor[HTML]{E7E7E7}{\color[HTML]{000000} FM}                    & \cellcolor[HTML]{E7E7E7}{\color[HTML]{000000}{$\left[\text{dB}\right]$}}              & \multicolumn{3}{c|}{\cellcolor[HTML]{E7E7E7}{\color[HTML]{000000} $10$}} \\ \cline{2-7} 
                        \cellcolor[HTML]{89B0FF}{\color[HTML]{002060}} & Margines shadowingu              & SM                    & {$\left[\text{dB}\right]$}              & \multicolumn{1}{c|}{$12.8$} & \multicolumn{1}{c|}{$15.2$} & \multicolumn{1}{c|}{$10$} \\ \cline{2-7} 
                       \multirow{-6}{*}{\rotatebox[origin=c]{90}{\cellcolor[HTML]{89B0FF}{\color[HTML]{002060} Propagacja}}} & \cellcolor[HTML]{E7E7E7}{\color[HTML]{000000} Strata implementacyjna}         & \cellcolor[HTML]{E7E7E7}{\color[HTML]{000000} IL}            & \cellcolor[HTML]{E7E7E7}{\color[HTML]{000000}{$\left[\text{dB}\right]$}}              & \multicolumn{1}{c|}{\cellcolor[HTML]{E7E7E7}{\color[HTML]{000000} $0$}}    & \multicolumn{1}{c|}{\cellcolor[HTML]{E7E7E7}{\color[HTML]{000000} $0$}}    & \multicolumn{1}{c|}{\cellcolor[HTML]{E7E7E7}{\color[HTML]{000000} $3$}} \\ \hline
    \end{tabular}}
\end{table}

\begin{table}[H]
\centering
\caption{Konfiguracja parametrów (IPR)
\cite{Tang}
}
\label{table:configuration_ris_design}
    \resizebox{0.49\textwidth}{!}{\begin{tabular}{|l|c|c|c|c|}
    \arrayrulecolor[HTML]{002060}\hline
                        \cellcolor[HTML]{002060}{\color[HTML]{FFFFFF}} & \cellcolor[HTML]{002060}{\color[HTML]{FFFFFF} Parametr}                           &  \cellcolor[HTML]{002060}{\color[HTML]{FFFFFF} Symbol}                     &  \cellcolor[HTML]{002060}{\color[HTML]{FFFFFF} Jednostka}                      & \cellcolor[HTML]{002060}{\color[HTML]{FFFFFF} Wartość} \\ \hline
                        \multirow{1}{*}{\cellcolor[HTML]{89B0FF}{\color[HTML]{002060}}} 
                        \cellcolor[HTML]{89B0FF}{\color[HTML]{002060}} & \cellcolor[HTML]{E7E7E7}{\color[HTML]{000000} Liczba}                   & \cellcolor[HTML]{E7E7E7}{\color[HTML]{000000} $K_\text{RIS}$}            & \cellcolor[HTML]{E7E7E7}{\color[HTML]{000000}\textbf{--}}                     &                                    \multicolumn{1}{c|}{\cellcolor[HTML]{E7E7E7}{\color[HTML]{000000}$15$}}      \\ \cline{2-5} 
                        \multirow{-6}{*}{\rotatebox[origin=c]{90}{\cellcolor[HTML]{89B0FF}{\color[HTML]{002060}}}} & {\color[HTML]{000000} L. elementów w kolumnie}         & {\color[HTML]{000000} $M_\text{RIS}$}            & {\color[HTML]{000000}{\textbf{--}}}              & \multicolumn{1}{c|}{{\color[HTML]{000000} $102$}} \\ \cline{2-5} 
                        \cellcolor[HTML]{89B0FF}{\color[HTML]{002060}} & \cellcolor[HTML]{E7E7E7}{\color[HTML]{000000} L. elementów w wierszu}        & \cellcolor[HTML]{E7E7E7}{\color[HTML]{000000} $N_\text{RIS}$}                    & \cellcolor[HTML]{E7E7E7}{\color[HTML]{000000}{\textbf{--}}}              & \multicolumn{1}{c|}{\cellcolor[HTML]{E7E7E7}{\color[HTML]{000000} $100$}}  \\ \cline{2-5} 
                        \cellcolor[HTML]{89B0FF}{\color[HTML]{002060}} & Szerokość kolumny             & $d_\text{m}$                    & {$\left[\text{m}\right]$}              & \multicolumn{1}{c|}{$0.01$} \\ \cline{2-5} 
                        \cellcolor[HTML]{89B0FF}{\color[HTML]{002060}} & \cellcolor[HTML]{E7E7E7}{\color[HTML]{000000} Szerokość wiersza}                & \cellcolor[HTML]{E7E7E7}{\color[HTML]{000000} $d_\text{n}$}                    & \cellcolor[HTML]{E7E7E7}{\color[HTML]{000000}{$\left[\text{m}\right]$}}              & \multicolumn{1}{c|}{\cellcolor[HTML]{E7E7E7}{\color[HTML]{000000} $0.01$}} \\ \cline{2-5} 
                        \cellcolor[HTML]{89B0FF}{\color[HTML]{002060}} & Współ. amp. syg. odbitego          & $A$                    & {\textbf{--}}              & \multicolumn{1}{c|}{$0.9$} \\ \cline{2-5}
                        \cellcolor[HTML]{89B0FF}{\color[HTML]{002060}} & \cellcolor[HTML]{E7E7E7}{\color[HTML]{000000} Wysokość zawieszenia} & \cellcolor[HTML]{E7E7E7}{\color[HTML]{000000} $h_\text{RIS}$}                    & \cellcolor[HTML]{E7E7E7}{\color[HTML]{000000} $\left[\text{m}\right]$}              & \multicolumn{1}{c|}{\cellcolor[HTML]{E7E7E7}{\color[HTML]{000000} $\left(18, 43\right)$}} \\ \cline{2-5}  
                        \multirow{1}{*}{\cellcolor[HTML]{89B0FF}{\color[HTML]{002060}}}  & {\color[HTML]{000000} Model tł. ścieżki} & {\color[HTML]{000000} \textbf{--}}                     & {\color[HTML]{000000} \textbf{--}}     & \multicolumn{1}{c|}{{\color[HTML]{000000} RIS FFBC \cite{Tang}}}\\ \hline
    \end{tabular}}
\end{table}

\section{Wyniki}
\label{section:results}
\begin{figure}[H]
\centering
\includegraphics[width=0.43\textwidth]{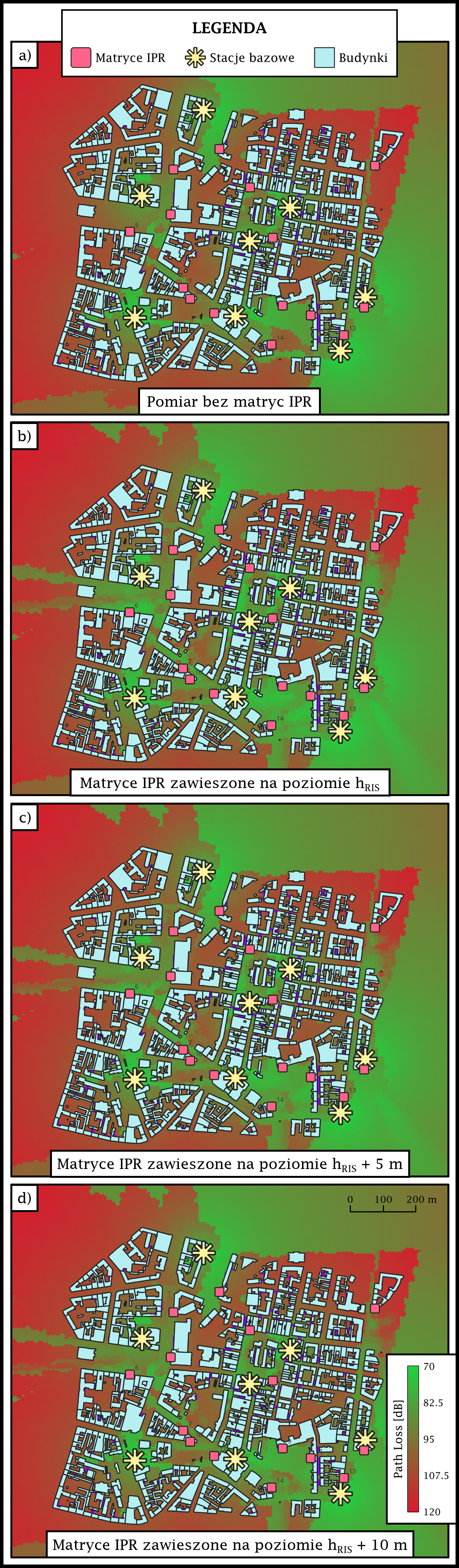}
\caption{Mapy rozkładu tłumienia ścieżki sygnału.}
\label{figure:pl_dist}
\end{figure}

Niniejsza sekcja prezentuje wyniki uzyskane podczas przeprowadzania badań symulacyjnych. Na Rys.~\ref{figure:pl_dist}.a zaprezentowano rozkład minimalnego tłumienia sygnału radiowego dla rozważanego obszaru gdy żadna z matryc nie jest brana pod uwagę. Kolejno Rys.~\ref{figure:pl_dist}.b\textbf{--}d prezentują te same wyniki, lecz po uwzględnieniu matryc IPR w przybliżaniu charakterystyki propagacyjnej oraz po zmanipulowaniu (lub też nie \textbf{--} Rys.~\ref{figure:pl_dist}.b) wysokości ich zawieszenia.

Porównując mapy zawarte w Rys.~\ref{figure:pl_dist} zauważyć można, iż wprowadzenie matryc IPR zredukowało tłumienie sygnału w niektórych rejonach badanego obszaru. Natomiast, łatwo również dostrzec, iż sama wysokość zawieszenia inteligentnych powierzchni szczególnie oddziałuje na prezentowaną charakterystykę propagacji sygnału radiowego. Analizując każdy przypadek z osobna (Rys.~\ref{figure:pl_dist}.b\textbf{--}d) można zobaczyć, iż zwiększając wysokość zawieszenia matryc IPR o pierwsze $5\text{ }\left[\text{m}\right]$ w południowej części obszaru zredukowano tłumienie sygnału radiowego. Jednakże kolejne podwyższenie punktu zaczepienia matryc odbijających (o $10\text{ }\left[\text{m}\right]$ względem wysokości domyślnej) niefortunnie pogarsza ten rozkład w porównaniu do Rys.~\ref{figure:pl_dist}.b\textbf{--}c.

Powyższe obserwacje zostały podsumowane w~Tab.~\ref{table:results}. Zawarte we wspomnianej tabeli wartości parametrów zostały uzyskane poprzez zastosowanie poniższych wzorów oznaczonych dalej jako Równ.~(\ref{equation:delta_h_ris})--(\ref{equation:delta_pl_avg}).

Pierwszym opisywanym parametrem jest średnia różnica zawieszenia matryc IPR między stanem obecnym a przyjętą wysokością domyślną $\left(\Delta h_\text{RIS}\right)$. Wysokość domyślna dla każdej matrycy jest równa w przybliżeniu wysokości budynku, na którym się znajduje. Wspomniana (średnia) różnica wysokości matryc odbijających może zostać opisana przez następującą formułę:
\begin{equation}
\label{equation:delta_h_ris}
    \Delta h_\text{RIS}=\frac{1}{K_\text{RIS}}\sum^{K_\text{RIS}-1}_{i=0}\Big(h_{\text{RIS},j}^\text{current}-h_{\text{RIS},j}^\text{default}\Big)
\end{equation}
gdzie $K_\text{RIS}$ to liczba powierzchni umieszczonych na badanym terenie, a $h_{\text{RIS},j}^\text{current}$ i $h_{\text{RIS},j}^\text{default}$ to odpowiednio wysokość $j$-tej $\left(j=0,\dots,K_\text{RIS}-1\right)$ matrycy IPR w chwili obecnej oraz jej wartość domyślna.

Kolejnym parametrem opisanym w Tab.~\ref{table:results} jest średnia różnica wysokości między stacjami bazowymi a matrycami IPR $\left(\Delta h_\text{BS--RIS}\right)$, które kierunkowały sygnał radiowy z tych stacji do poszczególnych punktów na mapie rozważanego obszaru. Wspomniany parametr opisuje następujące równanie:
\begin{equation}
\label{equation:delta_h_bs_ris}
    \Delta h_\text{BS--RIS}=\frac{\sum^{K_\text{BS}-1}_{i=0}\sum^{K_\text{RIS}-1}_{j=0}\chi_{i,j}\left(h_{\text{BS},i}-h_{\text{RIS},j}^\text{current}\right)}{\sum^{K_\text{BS}-1}_{i=0}\sum^{K_\text{RIS}}_{j=0}\chi_{i,j}}
\end{equation}
gdzie $K_\text{BS}$ to liczba komórek stacji bazowych znajdujących się w obrębie sieci mobilnej. Z kolei $h_{\text{BS},i}$ to wysokość zawieszenia anten $i$-tej komórki sieciowej $\left(i=0,\dots,K_\text{BS}-1\right)$. Natomiast $\chi_{i,j}$ to parametr binarny $\left(\chi_{i,j}\in\{0,1\}\right)$ przyjmujący wartość równą $1$, jeżeli $j$-ta  matryca IPR znajduje się w zasięgu {\it wzroku} $i$-tej komórki sieciowej przynajmniej w jednej z $L$ konfiguracji (dla niniejszego artykułu wyróżnić można $3$ przypadki \textbf{--} $\Delta h_{\text{RIS}}\in\left\{0, 5, 10\right\}\text{ }\left[\text{m}\right]$). W pozostałych przypadkach wartość $\chi_{i,j}=0$.

Najważniejszy z parametrów ewaluacyjnych, czyli średni zysk związany z redukcją tłumienia ścieżki sygnału radiowego $\left(\Delta \overline{\text{PL}}\right)$ został zaimplementowany w środowisku symulacyjnym w sposób następujący:
\begin{equation}
\label{equation:delta_pl_avg}
    \Delta \overline{\text{PL}}=\left[1-\frac{\sum^{Y-1}_{y=0}\sum^{X-1}_{x=0}\text{PL}_\text{RIS}^\text{min}\left(x,y\right)}{\sum^{Y-1}_{y=0}\sum^{X-1}_{x=0}\text{PL}_\text{NRIS}^\text{min}\left(x,y\right)}\right] \cdot 100\text{ }\%
\end{equation}
gdzie $x$ $\left(x=0,\dots,X-1\right)$ i $y$ $\left(y=0,\dots,Y-1\right)$ to koordynaty danego punktu znajdującego się w obrębie badanego obszaru miejskiego. Natomiast $\text{PL}_\text{RIS}^\text{min}\left(x,y\right)$ i~ $\text{PL}_\text{NRIS}^\text{min}\left(x,y\right)$ to odpowiednio wartości strat propagacyjnych sygnału radiowego (w $\left[\text{dB}\right]$) w punkcie $P\left(x,y\right)$ gdy w~ obliczeniach uwzględnione zostanie oddziaływanie matryc IPR na transmisję tegoż sygnału, oraz gdy to oddziaływanie zostanie pominięte. Wspomniane parametry opisano wewnątrz oprogramowania symulacyjnego tak, jak to przedstawiają poniższe wzory: 
\begin{align}
\label{equation:pl_ris_min}
    &\text{PL}_\text{RIS}^\text{min}\left(x,y\right)=\\ &\min\Bigl\{\text{PL}_{\text{LOS},i}\left(x,y\right);\text{}\text{PL}_{\text{NLOS},i}\left(x,y\right);\text{}\text{PL}_{\text{RIS},i,j}\left(x,y\right)\Bigr\} \nonumber
\end{align}
\begin{equation}
\label{equation:pl_no_ris_min}
    \text{PL}_\text{NRIS}^\text{min}\left(x,y\right)=\min\Bigl\{\text{PL}_{\text{LOS},i}\left(x,y\right);\text{}\text{PL}_{\text{NLOS},i}\left(x,y\right)\Bigr\}
\end{equation}
gdzie $\text{PL}_{\text{LOS},i}\left(x,y\right)$, $\text{PL}_{\text{LOS},i}\left(x,y\right)$ oraz $\text{PL}_{\text{RIS},i,j}\left(x,y\right)$ to wartości dla tłumienia ścieżki sygnału radiowego odbieranego w punkcie $P\left(x,y\right)$ od $i$-tej komórki sieciowej bezpośrednio $\left(\text{PL}_{\text{LOS},i}\right)$, pośrednio $\left(\text{PL}_{\text{NLOS},i}\right)$, tj. po osłabieniu jego mocy przez obiekty fizyczne (np. budynki) lub zjawiska atmosferyczne, oraz po odbiciu wiązki radiowej od $j$-tej matrycy IPR $\left(\text{PL}_{\text{RIS},i,j}\right)$. W obu tych przypadkach (Równ.~(\ref{equation:pl_ris_min}) i (\ref{equation:pl_no_ris_min})) zawsze wybierana jest korzystniejsza (minimalna) wartość tłumienia ścieżki.

Odnosząc się bezpośrednio do Tab.~~\ref{table:results} można zauważyć, iż po przekroczeniu progu, w którym to średnia różnica wysokości między stacjami bazowymi a matrycami IPR $\left(\Delta h_\text{BS--RIS}\right)$ przyjmuje wartości ujemne (powierzchnie odbijające wyższe od stacji bazowych w swoim zasięgu), zysk związany ze średnią mocą odbieranego sygnału radiowego na badanym obszarze $\left(\Delta\overline{\text{PL}}\right)$ zaczyna maleć o ok. $0.74\text{ }\%$ w porównaniu do najkorzystniejszego wariantu (matryce podniesione o $5\text{ }\left[\text{m}\right]$).

\begin{table}[H]
\centering
\caption{Wpływ wysokości zawieszenia matryc IPR na charakterystykę tłumienia ścieżki.}
\label{table:results}
\resizebox{0.49\textwidth}{!}{
\begin{tabular}{|c|c|c|c|}
\arrayrulecolor[HTML]{002060}
\hline
{\cellcolor[HTML]{002060}{\color[HTML]{FFFFFF} Parametr}} & \multicolumn{3}{c|}{\cellcolor[HTML]{002060}{\color[HTML]{FFFFFF} Wartość}} \\ \cline{2-4} 
\cellcolor[HTML]{E7E7E7}{\color[HTML]{000000} $\Delta h_\text{RIS}$} & \cellcolor[HTML]{E7E7E7}{\color[HTML]{000000} $0\text{ m}$} & \cellcolor[HTML]{E7E7E7}{\color[HTML]{000000} $5\text{ m}$} & \cellcolor[HTML]{E7E7E7}{\color[HTML]{000000} $10\text{ m}$}
\\ \hline 
\cellcolor[HTML]{FFFFFF}{\color[HTML]{000000} $\Delta h_\text{BS--RIS}$} & \cellcolor[HTML]{FFFFFF}{\color[HTML]{000000} $9.57\text{ m}$} & \cellcolor[HTML]{FFFFFF}{\color[HTML]{000000} $4.62\text{ m}$} & \cellcolor[HTML]{FFFFFF}{\color[HTML]{000000} $-0.96\text{ m}$}  \\ \hline
\cellcolor[HTML]{89B0FF}{\color[HTML]{002060} $\Delta \overline{\text{PL}}$} & \cellcolor[HTML]{89B0FF}{\color[HTML]{002060} $1.4\text{ }\%$} & \cellcolor[HTML]{89B0FF}{\color[HTML]{002060} $1.57\text{ }\%$} & \cellcolor[HTML]{89B0FF}{\color[HTML]{002060} $0.83\text{ }\%$}  \\ \hline
\end{tabular}}
\end{table}

\section{Wnioski}
\label{section:conclusions}
Badania przeprowadzone w kontekście wykorzystania inteligentnych matryc IPR w sieciach mobilnych wykazały istotny potencjał tego rozwiązania w poprawie jakości odbieranego sygnału radiowego w obszarach gęsto zabudowanych. Przy uwzględnieniu różnych konfiguracji wysokości zawieszenia powierzchni odbijających oraz ich rozmieszczenia, obserwowane zmiany w tłumieniu wiązki radiowej oraz jej propagacji mogą mieć znaczący wpływ na efektywność działania całej sieci RAN.

Analiza wyników przeprowadzonych badań pokazała, że zastosowanie matryc IPR pozwoliło na znaczną redukcję tłumienia wiązki w niektórych obszarach miasta. Dodatkowo manipulacja wysokością zawieszenia matryc miała istotny wpływ na charakterystykę propagacji sygnału radiowego. Zauważono, że zwiększając wysokość zawieszenia matryc o pewną wartość ($5\text{ }\left[\text{m}\right]$), można by poprawić warunki transmisyjne, jednakże dalsze podwyższenie może skutkować pogorszeniem jakości odbieranego sygnału w niektórych obszarach.

Ponadto analiza różnicy wysokości między stacjami bazowymi a matrycami IPR w rozważanych konfiguracjach symulacyjnych wykazała, że istnieje pewien próg, po którego przekroczeniu zysk związany z redukcją tłumienia sygnału może maleć. Jest to istotne spostrzeżenie przy projektowaniu i optymalizacji sieci, ponieważ sugeruje, że istnieje optymalna wysokość zawieszenia matryc IPR, która zapewnia najlepszą wydajność sieci mobilnej.

Podsumowując, wyniki badań potwierdziły, że inteligentne matryce IPR mają duży potencjał w poprawie jakości transmisji sygnału na obszarach miejskich. Jednakże ich efektywne wykorzystanie wymaga uwzględnienia wielu czynników, takich jak wysokość zawieszenia, rozmieszczenie oraz konfiguracja sieci RAN, co może być przedmiotem dalszych badań w celu optymalizacji działania (przyszłych) systemów bezprzewodowych.

\section*{Podziękowania}
Autorzy pragną podziękować prof. Margot Deruyck z~Uniwersytetu w Gandawie (IMEC) w Belgii za wsparcie niniejszej pracy poprzez udostępnienie oprogramowania GRAND.

Praca została zrealizowana w ramach projektu nr~ $2021/43/\text{B}/\text{ST}7/01365$ ufundowanego przez Narodowe Centrum Nauki (NCN) w Polsce.

Niniejszy artykuł jak i prace przedstawione w \cite{SamorzewskiKRiT2023, SamorzewskiSoftCOM2023, SamorzewskiGLOBECOM2023} zostały przygotowane we współpracy międzyuniwersyteckiej między Politechniką Poznańską a Uniwersytetem w~ Gandawie rozpoczętej w ramach programu grantowego Short-Term Scientific Mission (STSM) zorganizowanego przez akcję COST CA20120 INTERACT w roku 2023.

\end{multicols}
\end{document}